\documentclass[11pt]{article}
\pdfoutput = 1
\usepackage[utf8]{inputenc}

\usepackage[top=1in,bottom=1in,left=1in,right=1in]{geometry}
\makeatletter \g@addto@macro\@floatboxreset\centering \makeatother
\usepackage{color}
\definecolor{darkgreen}{rgb}{0,0.5,0}
\definecolor{darkblue}{rgb}{0,0,0.6}
\definecolor{purple}{rgb}{0.4,.2,0.7}
\usepackage[noEucal]{main}
\usepackage{aas_macros,empheq,fancybox,graphicx,multirow}
\usepackage{aas_macros}
\usepackage{ulem}
\usepackage{esint}
\usepackage{appendix}
\usepackage{mathabx}
\usepackage{tensor} 
\usepackage{footmisc}
\begin{document}

\thispagestyle{empty}

\begin{center}
    ~
    \vskip10mm

     {\LARGE  {\textsc{Soft Particles and Infinite-Dimensional Geometry}}}
    \vskip10mm
    
Daniel Kapec\\
    \vskip1em

    {\it
        Center of Mathematical Sciences and Applications, Harvard University, Cambridge, Massachusetts 02138, USA\\ \vskip1mm
         \vskip1mm
    }
    {\it
        Center for the Fundamental Laws of Nature, Harvard University, Cambridge, Massachusetts 02138, USA\\ \vskip1mm
         \vskip1mm
    }
    \vskip5mm
    \tt{danielkapec@fas.harvard.edu}
\end{center}
\vspace{10mm}

\begin{abstract}
\noindent
In the sigma model, soft insertions of moduli scalars enact parallel transport of $S$-matrix elements about the finite-dimensional moduli space of vacua, and the antisymmetric double-soft theorem calculates the curvature of the vacuum manifold. We explore the analogs of these statements in gauge theory and gravity in asymptotically flat spacetimes, where the relevant moduli spaces are infinite-dimensional. These models have spaces of vacua parameterized by (trivial) flat connections on the celestial sphere, and soft insertions of photons, gluons, and gravitons parallel transport $S$-matrix elements about these infinite-dimensional manifolds. We argue that the antisymmetric double-soft gluon theorem in $d+2$ bulk dimensions computes the curvature of a connection on the infinite-dimensional space Map$(S^d,G)/G$, where $G$ is the global part of the gauge group. The analogous metrics in abelian gauge theory and gravity are flat, as indicated by the vanishing of the antisymmetric double-soft theorems in those models. In other words, Feynman diagram calculations not only know about the vacuum manifold of Yang-Mills theory, they can also be used to compute its curvature. The results have interesting implications for flat space holography.

\end{abstract}
\pagebreak

\setcounter{tocdepth}{2}
{\hypersetup{linkcolor=black}
\small
\tableofcontents
}

\section{Introduction}
This paper is about an  apparent ambiguity in the definition of the Yang-Mills $S$-matrix. In the Coulomb phase, these $S$-matrix elements can be computed in perturbation theory and are unambiguous for finite, nonzero momenta. However, the multi-soft limits of the $S$-matrix, in which the wavelengths of two or more gluons approach infinity, depend on the order of limits and are therefore ambiguous. This ambiguity is quantified by the  commutator of consecutive soft limits, also known as the antisymmetric double-soft theorem. These antisymmetric soft limits are universal (at tree-level) but non-vanishing for any non-abelian gauge group $G$:
\begin{equation} \label{eq:Ambiguity}
 \left[\lim_{q_I\to 0},\lim_{q_J\to 0} \right]A_{n+2}^{K_1\cdots K_n; IJ,ab}\neq 0 \; .
\end{equation}
There is some confusion regarding the interpretation of this fact. Strictly speaking, the soft limit requires one to consider the boundary of the domain upon which the $S$-matrix is defined, and one could phrase \eqref{eq:Ambiguity} as an ambiguity in the extension of the $S$-matrix to the boundary of kinematic space. Although one is tempted to look for the ``right definition'' of the multi-soft limits, compactifications of this sort arising in moduli problems are often non-canonical\footnote{We thank Andrew Strominger for  emphasizing this point.} and any prescription is bound to appear somewhat arbitrary.
Although this is really just a question about the four-dimensional $S$-matrix, the issue takes on added importance in the celestial CFT (CCFT) approach to flat space holography, where soft gluons are related to non-abelian conserved currents in a lower-dimensional boundary dual. Due to \eqref{eq:Ambiguity}, the correlation functions of these currents appear ambiguous.

In this paper we will advocate for an (infinite-dimensional) geometric interpretation of \eqref{eq:Ambiguity} which combines observations on the vacuum structure of gauge theory \cite{Strominger:2013jfa,Strominger:2013lka,Kapec:2014opa,Kapec:2014zla} with recent geometric results on soft limits in the sigma model \cite{Cheung:2021yog,Kapec:2022axw}. 
In a word, \eqref{eq:Ambiguity} is true because the amplitude $A_{n+2}^{K_1\cdots K_n; IJ,ab}$ is actually a section of a bundle over an infinite-dimensional space of vacua $\mathcal{M}$ labeled by flat (trivial) $G-$connections $\mathcal{C}_a(x)$ on the celestial sphere. Insertions of soft gluons in the $S$-matrix enact parallel transport about this space, and \eqref{eq:Ambiguity} simply computes the  holonomy around an infinitesimal closed curve, also known as the Riemann curvature $R(X,Y)Z=[\nabla_X,\nabla_Y]Z-\nabla_{[X,Y]}Z$  
\begin{equation}\label{eq:YMcurv}
 \left[\lim_{q_I(x)\to 0}\omega,\lim_{q_J(y)\to 0}\omega' \right]A_{n+2}^{K_1\cdots K_n; IJ,ab}=   R\left(\frac{\delta }{\delta \widetilde{\mathcal{C}^I_a}(x)}, \frac{\delta }{\delta \widetilde{\mathcal{C}^J_b}(y)} \right)A_n^{K_1\cdots K_n} \;.
\end{equation}
In this formula, the (shadow-transformed) flat connections $\widetilde{\mathcal{C}_a}(x)$ are local coordinates on $\mathcal{M}$, $\frac{\delta }{\delta \widetilde{\mathcal{C}^I_a}(x)}$ is an associated vector field, and the fact that the antisymmetric double-soft limit does not vanish simply means that the vacuum manifold for Yang-Mills theory in asymptotically flat space is curved.
In other words, there is no ``right definition'' for multi-soft limits in non-abelian gauge theory, just as there is no unique parallel transport between two points in a curved space. The result is path dependent, and \eqref{eq:Ambiguity} is just an infinite-dimensional manifestation of that geometric fact.
Note that in this language, the vanishing of the antisymmetric double-soft limit in gravity is a statement about flatness of the space of supertranslation vacua, and a similar statement holds for abelian gauge theory.

The claim that there exists an infinite-dimensional space of vacua for gauge theory in asymptotically flat space is not new. In fact, this observation \cite{Strominger:2013lka}, and its analog in gravity \cite{Strominger:2013jfa}, was the genesis of the ``celestial CFT'' program which seeks to construct a boundary dual for asymptotically flat gravity. The existence of these vacua is a subtle boundary effect which only holds for gauge theories in the Coulomb phase on non-compact spaces, but the interpretation of \eqref{eq:Ambiguity} advanced in this paper makes it clear that the moduli space of vacua is detectable using standard perturbative calculations: the formula \eqref{eq:Ambiguity} can be derived using standard Feynman rules with no reference to asymptotic symmetries, boundary conditions, or gauge transformations with non-compact support. Perturbative $S$-matrix calculations in Yang-Mills theory (which are by definition performed in the Coulomb phase, with long-range gauge fields) know about the infinite-dimensional space of vacua. In fact they know more: they can calculate its curvature.

To put these statements in context, note that the antisymmetric double-soft limit has historically been a discovery mechanism for hidden symmetries of the $S$-matrix (see for instance \cite{Arkani-Hamed:2008owk} for the case of $E_{7(7)}$ in $\mathcal{N}=8$ supergravity in four dimensions). The gauge theory examples discussed in this work are really just infinite-dimensional examples of this phenomena. Indeed, these statements are all simpler and more familiar when the moduli space of vacua $M$ is finite-dimensional. In this case, the vacuum
is determined by boundary conditions (vacuum expectation values, or vevs) at spatial infinity ($i^0$) for gapless local fields $\langle \Phi^I \rangle_{i^0} =v^I$, and the dynamics of the long-wavelength fluctuations about these vevs is described by a sigma model with target space $M$. The single soft insertion of a moduli scalar defines an operator whose matrix element is a derivative on the moduli space \cite{Cheung:2021yog,Kapec:2022axw}
\begin{equation}\label{eq:NLSMsoft}
\langle S_I(x) \mathcal{O}_1 \cdots \mathcal{O}_n \rangle_v  =  \nabla_I\; \langle \mathcal{O}_1 \cdots \mathcal{O}_n \rangle_{v} \; .
\end{equation}
Here, $S_I(x)$ is a soft moduli scalar, the $\mathcal{O}_i$ represent hard particles, and the subscript $\langle \cdot \rangle_{v}$ indicates that the $S$-matrix element is calculated with the asymptotic boundary condition $\langle \Phi^I \rangle_{i^0} =v^I$. The physical interpretation of this formula is simple: the zero mode of the moduli scalar is just the vev, so exciting this mode infinitesimally shifts the boundary condition. 
In the celestial CFT formalism, the existence of this operator is equivalent to the existence of a marginal deformation in CCFT \cite{Kapec:2022axw}: $S_I(x)$ indicates a continuous, finite-dimensional family of celestial CFTs labeled by boundary conditions and related by marginal perturbations (i.e., constant background fields for the marginal operator). 
Multi-soft limits in the sigma model do not commute, and the antisymmetric double-soft theorem measures the infinitesimal holonomy around a closed loop  in the space of vacua  \cite{Cheung:2021yog}
\begin{equation}\label{eq:NLSMintro}
 \left[\lim_{q_I\to 0},\lim_{q_J\to 0} \right]A_{n+2}^{K_1\cdots K_n IJ}=   \left[\nabla^{I}, \nabla^{J} \right]A_n^{K_1\cdots K_n}= \sum_{i=1}^n R^{IJK_i}_{\quad \;K}A_n^{K_1\cdots K\cdots K_n} \; .
\end{equation}
The analog of \eqref{eq:NLSMsoft} in gauge theory is the soft gluon theorem
\begin{equation}\label{eq:YMsoftIntro}
\langle S_a^I(x) \mathcal{O}_1 \cdots \mathcal{O}_n \rangle_{\mathcal{C}=0}  = ig_{\text{YM}} \sum_{k=1}^n\frac{p_k \cdot \varepsilon_a(x)}{p_k \cdot \hat{q}(x)}T_k^I\; \langle \mathcal{O}_1 \cdots \mathcal{O}_n \rangle_{\mathcal{C}=0} \; ,
\end{equation}
where $S_a^I(x)$ is a soft gluon with  transverse polarization index $a$, color index $I$, and momentum in the direction $\hat{q}(x)$, and the subscript $\langle \cdot \rangle_{\mathcal{C}}$ indicates that the $S$-matrix element is calculated with the asymptotic boundary condition $\langle A_a\rangle=\mathcal{C}_a(x)$. We would like to interpret this formula by analogy with the the sigma model: the zero mode of the gluon shifts the boundary condition, so \eqref{eq:YMsoftIntro} is really a functional derivative on the space of flat $G-$connections on the celestial sphere. It is by now understood that $S_a^I(x)$ is related to the non-abelian current in the CCFT dual \cite{He:2015zea,Kapec:2017gsg,Kapec:2021eug}, and in this language the preceding statement seems very similar to the definition of the current correlator in the presence of a background source (in this case, a flat connection)
\begin{equation}
    \langle J^I_a(x) \mathcal{O}_1\cdots \mathcal{O}_n \rangle_{\mathcal{C}=0} = \frac{\delta}{\delta \mathcal{C}_I^a(x)}\langle \mathcal{O}_1\cdots \mathcal{O}_n  \rangle_{\mathcal{C}}\big{|}_{\mathcal{C}=0} \; .
\end{equation}
This formula obviously resembles \eqref{eq:NLSMsoft}. Proceeding with the analogy, we will be led to interpret the antisymmetric double-soft gluon theorem as an infinite-dimensional version of \eqref{eq:NLSMintro}. 
The fact that \eqref{eq:NLSMsoft} and \eqref{eq:NLSMintro} have no momentum dependence (position dependence in CCFT) while \eqref{eq:Ambiguity} and \eqref{eq:YMsoftIntro} do is a signal that the moduli space in gauge theory is infinite-dimensional, i.e., a space of functions.
This is related to the position dependence of the boundary conditions $\langle A_a\rangle = \mathcal{C}_a(x)$ which is absent in the sigma model.

The organization of this paper is as follows. In section \ref{sec:Kinematics} we review the celestial CFT formalism. Section \ref{sec:sigma} reviews the finite-dimensional case of the sigma model and its relation to marginal deformations of CCFT. Sections \ref{sec:Abelian} and \ref{sec:gravity} treat the infinite-dimensional flat examples of abelian gauge theory and gravity. Section \ref{sec:Yang-Mills} treats the non-abelian case, and section \ref{sec:Discussion} concludes with a discussion of the implications for celestial CFT.

\section{Kinematics and celestial CFT$_d$}\label{sec:Kinematics}
The results of this paper are logically independent of the celestial CFT formalism, but they have interesting implications for CCFT and are most easily phrased in that language.

We will work on $\mathbb{R}^{d+1,1}$. The Lorentz group $SO(d+1,1)$ is isomorphic to the $d$-dimensional Euclidean conformal group. If we parameterize a generic massless on shell momentum as
\begin{equation}\label{eq:mompar}
q^\mu(\omega,x) = \omega {\hat q}^\mu ( x ) \; , \qquad {\hat q}^\mu(x) = \frac{1}{2} \left( 1 + x^2 , 2 x^a , 1 - x^2 \right) ,
\end{equation}
then Lorentz transformations act as global conformal transformations on the $x^a$ coordinates (the transverse cuts of the null momentum cone, or equivalently the transverse cuts of null infinity). The $d$ independent transverse polarization vectors are
\begin{equation}\label{eq:gluonpol}
\varepsilon_a^\mu(x) \equiv \partial_a \hat{q}^\mu(x) =  ( x_a , \delta_a^b, - x_a  ) \; , 
\end{equation}
and they satisfy 
\begin{equation}
n \cdot \varepsilon_a(x) = 0 \; , \quad {\hat q}(x_i) \cdot \varepsilon_a(x_j) = ( x_{ij} )_a \; , \quad \varepsilon_a(x_i) \cdot \varepsilon_b(x_j) = \delta_{ab}  \; , \quad -2\hat{q}(x)\cdot \hat{q}(y)=(x-y)^2 \; ,
\end{equation}
with $n=\frac12 (1,0^a,-1)$.
Similarly, the transverse traceless polarization tensor for gravitons is
\begin{equation}
\label{eq:gravpol}
\varepsilon^{\mu\nu}_{ab}(x) \equiv\frac{1}{2}  [ \varepsilon_a^{\mu}(x) \varepsilon^{\nu}_b(x) + \varepsilon_a^\nu(x) \varepsilon^\mu_b(x)  ] - \frac{1}{d} \delta_{ab} \Pi^{\mu\nu}(x)  \;, 
\end{equation}
where
\begin{equation}
\Pi^{\mu\nu}(x) \equiv \varepsilon^\mu_a(x) \varepsilon^{a,\nu} (x) = \eta^{\mu\nu} + 2 n^{\mu} {\hat q}^{\nu}(x) + 2 n^{\nu} {\hat q}^{\mu}(x) \; . 
\end{equation}
The operators that create single particle states live on the null cone
\begin{equation}\label{eq:OperatorDef}
\mathcal{O}_i(\omega_i,x_i) \equiv a_i^\text{out}(q(\omega_i,x_i)) \theta(\omega_i) +  \bar{a}_i^{\text{in}\dagger}(-q(\omega_i,x_i)) \theta(-\omega_i) \;,
\end{equation}
and the $S$-matrix amplitudes can be written in the suggestive form
\begin{equation}\label{eq:ConformalRep}
\mathcal{A}_n =  \langle \mathcal{O}_1(\omega_1,x_1) \cdots \mathcal{O}_n(\omega_n,x_n) \rangle \; . 
\end{equation}
The amplitude \eqref{eq:ConformalRep} does not transform like a correlator of conformal primaries in CFT$_d$ since momentum eigenstates are not boost eigenstates.  For massless particles, this is fixed by performing a Mellin transform
\begin{equation}
    \widehat{\mathcal{O}}(\Delta,x) =\int_\mathcal{C} d\omega \,\omega^{\Delta-1}\mathcal{O}(\omega,x) \; 
\end{equation}
for some contour $\mathcal{C}$ and scaling dimension $\Delta$.
We will be interested in the ``residue operators'' arising from Mellin transforms with integer dimensions and compact contours surrounding the origin   
\begin{equation}\label{eq:compactcontour}
\widehat{\mathcal{O}}(n,x) =\oint \frac{d\omega}{2\pi i }  \,\omega^{n-1}\mathcal{O}(\omega,x) \; . 
\end{equation}
These operators isolate terms in the soft expansion of scattering amplitudes and are therefore infrared operators: their matrix elements do not depend on the high-energy behavior of the amplitudes. For instance, the leading soft moduli operator in the sigma model  has $\Delta=0$ and spin-$0$
\begin{equation}\label{eq:SIintro}
    S^I(x)= \oint \frac{d\omega}{2\pi i}\omega^{-1}\mathcal{O}^I(\omega,x) \; ,
\end{equation}
while the leading soft gluon operator has $\Delta=1$ and spin-$1$
\begin{equation}\label{eq:SIaintro}
    S^I_a(x)= \oint \frac{d\omega}{2\pi i}\mathcal{O}^I_a(\omega,x) \; .
\end{equation}
There is an alternative definition for these operators in terms of ``conformally soft limits'' \cite{Donnay:2018neh,Pate:2019mfs}
\begin{equation}\label{eq:ConfSoft}
    \lim_{\Delta \to n}(\Delta-n)\int_0^\infty d\omega \,\omega^{\Delta-1}\mathcal{O}(\omega,x) \; .
\end{equation}
Roughly speaking, the matrix elements agree since the Mellin transform of a Laurent expansion in $\omega$ produces poles at integer values of $\Delta$ with residues fixed by the Laurent coefficients.\footnote{ The definitions \eqref{eq:SIintro}-\eqref{eq:SIaintro} are more intuitive, but when there are branch cuts in the momentum-space amplitudes \eqref{eq:ConfSoft} is the appropriate generalization.}

The operators $S^I(x)$ and $S^I_a(x)$ are related to marginal deformations and conserved currents of CCFT through the shadow transform \cite{Kapec:2017gsg,Kapec:2022axw}, which takes the general form
\begin{equation}
    \widetilde{ \mathcal{O}}(x) \equiv \int d^d y \frac{ 1 }{ [ ( x - y )^2 ]^{d-\Delta} } \mathcal{R}  ( \mathcal{I}(x-y) ) \cdot \mathcal{O}(y) \; ,
\end{equation}
where $\mathcal{R}(\mathcal{I})$ is the appropriate representation of the conformally covariant tensor. For scalar operators $\mathcal{R}(\mathcal{I})=1$, while for vectors $\mathcal{I}_{ab}(x)=\delta_{ab}-2\frac{x_ax_b}{x^2}$. Importantly, the shadow transform squares to a multiple of the identity
\begin{equation}\label{eq:ShadowSquared}
\widetilde{ \widetilde{ \mathcal{O}}} (x) = c_{\Delta,s} \mathcal{O}(x)\;  , \qquad c_{\Delta,s} = \frac{\pi^d (\Delta-1)(d-\Delta-1)\Gamma(\frac{d}{2}-\Delta)\Gamma(\Delta-\frac{d}{2})}{(\Delta-1+s)(d-\Delta-1+s)\Gamma(\Delta)\Gamma(d-\Delta)} \; .  
\end{equation}

\section{Sigma model}\label{sec:sigma}
When a quantum field theory in asymptotically flat space has a continuous moduli space of vacua, the boundary conditions at spatial infinity $\langle \Phi \rangle_{i^0}=v^I$ provide local coordinates for the vacuum manifold $M$. The dynamics of the low energy, long-wavelength fluctuations about these vevs is described by a sigma model with target space $M$ and the minimal action is
\begin{equation}\label{eq:SigmaAction}
    S=\frac12\int d^{d+2}x \; G_{IJ}(\Phi)\partial_\mu \Phi^I\partial^\mu \Phi^J \; .
\end{equation}
$G_{IJ}(\Phi)$ is the Riemannian metric on $M$ with curvature $R_{IJKL}$. $S$-matrix elements in this model carry an extra label $\langle \cdot \rangle_{v}$ indicating that they are computed with the boundary condition $\langle \Phi \rangle_{i^0}=v$. In order to perform perturbative calculations, we expand the fields about their vevs $\Phi^I=v^I+\phi^I$ and integrate over the normalizable modes. Physical observables are expressed in terms of the intrinsic geometry of $M$. For instance, the soft limit of a moduli scalar takes the form \cite{Cheung:2021yog}
\begin{equation}\label{eq:singlesoft thm}
    \lim_{\omega\to 0} \;\langle \mathcal{O}_I(\omega,x)\mathcal{O}_1(\omega_1,x_1)\cdots \mathcal{O}_n(\omega_n,x_n)\rangle_{v} = \nabla_I \; \langle \mathcal{O}_1(\omega_1,x_1)\cdots \mathcal{O}_n(\omega_n,x_n)\rangle_v \; .
\end{equation}
This formula has a simple interpretation: since the zero mode of the scalar is the vev, exciting the zero mode simply shifts the boundary condition infinitesimally and parallel transports the $S$-matrix about the space of vacua. The leading soft moduli operator
\begin{equation}\label{eq:SoftOP}
    S_I(x)\equiv \oint \frac{d\omega}{2\pi i }\omega^{-1}\mathcal{O}_I(\omega,x) \; 
\end{equation}
has momentum independent matrix elements (position independent in the language of CCFT)
\begin{equation}\label{eq:singlesoft}
\langle S_I(x) \mathcal{O}_1 \cdots \mathcal{O}_n \rangle_v  =  \nabla_I\; \langle \mathcal{O}_1 \cdots \mathcal{O}_n \rangle_{v} \; .
\end{equation}
Since $S_I(x)$ has $\Delta=0$, it would arise naturally as the shadow transform of a $\Delta=d$ scalar operator, i.e., a \textit{marginal deformation}
\begin{equation}
    S_I = \int d^dy \; M_I(y) \; .
\end{equation}
As described in \cite{Kapec:2022axw}, if we view $S_I$ as the integrated deformation one adds to the action of the CCFT when deforming by an exactly marginal operator, then \eqref{eq:singlesoft} is simply the formula for first-order conformal perturbation theory in CCFT. A finite deformation (coherent state of soft scalars) turns on a background source for the marginal operator
\begin{align}\label{eq:transport}
    \langle \mathcal{O}_1 \cdots \mathcal{O}_n \rangle_{v-\lambda}
    &=\langle \mathcal{O}_1 \cdots \mathcal{O}_n \exp\left[-\lambda^I S_I\right] \rangle_{v}\\
    &\equiv\langle \mathcal{O}_1 \cdots \mathcal{O}_n \exp\left[-\lambda^I \int d^dxM_I(x)\right] \rangle_{v} \; . \notag
\end{align}
When the moduli space is curved and we deform in multiple directions, the resulting observable becomes path dependent. Taking this path to be an infinitesimal closed loop computes the curvature \cite{Cheung:2021yog}
\begin{equation}\label{eq:DoubleSoftComm}
 \left[\lim_{q_I\to 0},\lim_{q_J\to 0} \right]A_{n+2}^{K_1\cdots K_n IJ}=   \left[\nabla^{I}, \nabla^{J} \right]A_n^{K_1\cdots K_n}  \; .
\end{equation}
In flat space holography, this is interpreted as the curvature of the conformal manifold of celestial CFT \cite{Kapec:2022axw}.

\section{Abelian gauge theory}\label{sec:Abelian}
Finite energy boundary conditions for abelian gauge theory in asymptotically flat space allow for a leading trivial (pure gauge) flat connection at infinity \cite{Strominger:2013lka,Kapec:2014zla}
\begin{equation}\label{eq:abelianBC}
    \langle A_a\rangle=\mathcal{C}_a(x) \; , \qquad \partial_{[a}\mathcal{C}_{b]}(x)=0 \; .
\end{equation}
This notation indicates that, in an asymptotic expansion about null infinity in flat coordinates \cite{He:2019jjk} ${ds^2=-dudr+r^2 \delta_{ab}dx^adx^b}$, the transverse components of the gauge field have large-$r$ asymptotics ${A_a(u,r,x)\sim \mathcal{C}_a(x) + O(r^{-1})}$, with $\mathcal{C}_a(x)$ a (trivial) flat connection independent of the retarded time $u$. The radiative degrees of freedom in the electromagnetic field appear in the subleading terms in the large-$r$ expansion.\footnote{ In four bulk dimensions the radiative data is not subleading, but the $O(r^0)$ term can be decomposed into a radiative term that depends on $u$ and a non-normalizable boundary mode $\mathcal{C}_a(x)$ that obeys \eqref{eq:abelianBC}.} Since this degree of freedom $\mathcal{C}_a(x)$ contains no $u$-dependence, it is effectively non-dynamical at the classical level. Nonetheless, it plays an important role in the infrared dynamics of the quantum theory.

The boundary condition \eqref{eq:abelianBC} is invariant under global $U(1)$ gauge transformations, but it spontaneously breaks all other gauge transformations with non-compact support.\footnote{The formulation of the model in terms of the electric field variables can obscure magnetic symmetries, which are most simply discussed in terms of the dual magnetic gauge potential $\widetilde{A}$. The symmetry structure in the presence of magnetic sectors deserves further investigation (and is even more important in the non-abelian case) but will not be the focus of this paper. For related discussions see \cite{Nande:2017dba,Kapec:2021eug}. }  The space of boundary conditions is simply the orbit of the zero connection $\mathcal{C}_a=0$ under the action of $\mathcal{G}=$Map$(S^d,U(1))$. The moduli space is therefore $\mathcal{M}=\mathcal{G}/U(1)$ since the global $U(1)$ is unbroken on the vacuum.

In the quantum theory, $S$-matrix elements $\langle \cdot \rangle_{\mathcal{C}}$ carry an extra label corresponding to the boundary condition \eqref{eq:abelianBC}. Since that boundary condition spontaneously breaks $\mathcal{G}\to U(1)$, there is an associated Goldstone boson $C_a(x)$ for the coset. 
Under large gauge transformations 
\begin{equation}
    C_{a}(x)\to  C_{a}(x) + \partial_a \varepsilon(x) \; , 
\end{equation}
so this degree of freedom indeed behaves like a Goldstone boson\footnote{This use of the term ``Goldstone boson'' simply refers to a flat direction in field space protected by a spontaneously broken symmetry. There is no massless scalar particle in the bulk spectrum, but there is a sense in which the photon is a generalized Goldstone boson associated to the pattern of symmetry breaking $\text{Map}(S^d,U(1))\to U(1)$. } for the infinite-dimensional group $\mathcal{G}$. Any infrared regulator explicitly breaks the $\mathcal{G}$ symmetry, so in the presence of an infrared regulator $C_a(x)$ is really a \textit{pseudo}-Goldstone boson.

Perturbative $S$-matrix calculations are usually performed with the boundary condition $\mathcal{C}=0$, and the soft photon theorem says
\begin{equation}
\langle \mathcal{O}_a ( \omega , x ) \mathcal{O}_1 \cdots \mathcal{O}_n \rangle_{\mathcal{C}=0} ~~ \stackrel{\o\to0}{\longrightarrow} ~~ \frac{1}{\o} e \sum_{k=1}^n Q_k \frac{ {\hat p}_k \cdot \varepsilon_a(x) }{ {\hat p}_k \cdot {\hat q}(x)  } \langle \mathcal{O}_1 \cdots \mathcal{O}_n \rangle_{\mathcal{C}=0}  \;  .
\end{equation}
The leading soft photon operator 
\begin{equation}
    S_a(x)= \frac{1}{e}\oint \frac{d\omega}{2\pi i}\mathcal{O}_a(\omega,x) \; 
\end{equation}
has matrix elements 
\begin{equation}\label{eq:abelianSoft}
\langle S_a(x) \mathcal{O}_1 \cdots \mathcal{O}_n \rangle_{\mathcal{C}=0} = \mathcal{J}_a (x) \langle \mathcal{O}_1 \cdots \mathcal{O}_n  \rangle_{\mathcal{C}=0} \;  ,  \qquad \mathcal{J}_a (x) = \partial_a \sum_{k=1}^n Q_k \log [ - {\hat p}_k \cdot {\hat q}(x) ] \;  ,
\end{equation}
and a straightforward computation shows that the shadow transform
\begin{equation}\label{eq:AbelianCurrent}
    J_a(x)=\frac{1}{2c_{1,1}}\int d^dy \frac{\mathcal{I}_{ab}(x-y)}{[(x-y)^2]^{d-1}}S_a(y)
    \equiv \frac{1}{2c_{1,1}}\widetilde{S}_a(x)
\end{equation} obeys the Ward identity for a conserved current\footnote{$c_{1,1}$ vanishes in odd dimensions, but the shadow integral in \eqref{eq:AbelianCurrent} vanishes as well. A finite result is obtained by evaluating the RHS of \eqref{eq:AbelianCurrent} for general $\Delta$ and then taking the $\Delta \to 1$ limit. For the special case $d=2$, $c_{1,1}=4\pi^2$ rather than $\pi^2$ due to special properties of the shadow transform in two dimensions \cite{Kapec:2021eug}.}
\begin{equation}\label{eq:AbelianWI}
\langle \partial^a J_a(x) \mathcal{O}_1 \cdots \mathcal{O}_n  \rangle_{\mathcal{C}=0} =  \sum_{k=1}^n Q_k \delta^{(d)}(x-x_k)  \langle  \mathcal{O}_1 \cdots \mathcal{O}_n \rangle_{\mathcal{C}=0} \;  .
\end{equation}
Note that the gauge theory soft factor $\frac{p_k\cdot \varepsilon}{p_k\cdot q}$ is $O(\omega^{-1})$ in the soft particle energy, but independent of the hard particle energy. For this reason, the $d$-dimensional conserved currents in CCFT obey the correct Euclidean Ward identity independently of whether or not the other operators in the correlator are Mellin transformed (the same is true of the NLSM soft theorem, which does not depend on the hard particle energies). In other words, the $d$-dimensional symmetry structure of $d+2$-dimensional gauge theory is evident even in the standard plane-wave basis.

In the presence of an IR regulator $\mu$ and a hard-soft separation scale $\Lambda$, the soft factorization theorems \cite{Bassetto:1983mvz,Feige:2014wja,Schwartz:2014sze} say
\begin{equation}\label{eq:AbelianExchange}
    \langle \mathcal{O}_1 \cdots \mathcal{O}_n \rangle_{\mathcal{C}=0}^\mu  = e^{ - \Gamma } \langle \mathcal{O}_1 \cdots \mathcal{O}_n \rangle_{\mathcal{C}=0}^\Lambda \; , \qquad  \Gamma=\alpha \int  \frac{d^d x}{(2\pi)^d} [\mathcal{J}_a(x)]^2 \; , \qquad \alpha= \frac{e^2}{8 \pi}\int^{\Lambda}_\mu \omega^{d-3}d\omega \; .
\end{equation}
These formulas result from re-summing virtual soft exchange in the range $\mu<\omega<\Lambda$ to all orders in perturbation theory. The fact that $\Gamma$ is expressible as a $d$-dimensional quantity is another indication that the soft dynamics is effectively $d$-dimensional. 

Following the analogy with the sigma model, we would like to interpret \eqref{eq:abelianSoft} as a derivative in the space of vacua, in this case the space of flat $U(1)$ connections on the celestial sphere. In other words, we expect a formula of the schematic form
\begin{equation}
    \langle S_a(x) \mathcal{O}_1\cdots \mathcal{O}_n \rangle \stackrel{?}{\sim} \frac{\delta}{\delta \mathcal{C}^a(x)}\langle \mathcal{O}_1\cdots \mathcal{O}_n  \rangle \; .
\end{equation}
Because the infrared sector of abelian gauge theory is exactly solvable, we can actually derive and verify these abstract statements. It was shown in \cite{Kapec:2021eug} that the (Euclidean) soft effective  action for the Goldstone mode and the soft photon mode in abelian gauge theory is given by
\begin{equation}\label{eq:SoftAction}
S_{\text{soft}}  = \frac{\alpha}{(2\pi)^d} \int d^d x \, S^a(x) S_a(x)  - \frac{i}{2c_{1,1}} \int d^d x \, \widetilde{C}_a (x) [ S^a (x) - \mathcal{J}^a(x) ] \; .
\end{equation}
Here $\widetilde{C}$ is the shadow transform of the Goldstone mode. This action reproduces all known facts about the infrared sector of abelian gauge theory, including soft theorems and exponentiated soft exchange \eqref{eq:AbelianExchange}
\begin{equation}\label{eq:IRanswer}
\langle S_{a_1}(y_1) \cdots S_{a_m}(y_m) \rangle_{\mathcal{C}=0} =   \exp \left[ - \frac{\alpha}{(2\pi)^d} \int d^d x \, \mathcal{J}^a(x) \mathcal{J}_a(x)   \right] \mathcal{J}_{a_1}(y_1) \cdots \mathcal{J}_{a_m}(y_m) \; .
\end{equation}
When the boundary condition for the asymptotic connection is nonzero, this action becomes
\begin{equation}
S_{\text{soft}}[\mathcal{C}(x)]  = \frac{\alpha}{(2\pi)^d} \int d^d x \, S^a(x) S_a(x)  - \frac{i}{2c_{1,1}} \int d^d x \, \widetilde{C}_a (x) [ S^a (x) - \mathcal{J}^a(x)] +\widetilde{\mathcal{C}}_a(x)S^a(x) \; .
\end{equation}
The path integral with the background field can also be done exactly. It is simply
\begin{equation}\label{eq:SoftInt}
    Z[\mathcal{C}]= \exp \left[ - \frac{\alpha}{(2\pi)^d} \int d^d x \, \mathcal{J}^a(x) \mathcal{J}_a(x)   \right]\exp\left[\frac{i}{2c_{1,1}}\int d^d x \, \widetilde{\mathcal{C}}_a(x)\mathcal{J}^a(x)\right] \; .
\end{equation}
Comparing with \eqref{eq:IRanswer}, this formula makes it clear that\footnote{In this and the following section, we use the notation $\frac{\delta}{\delta C(x)}$ to denote the functional partial derivative, which is also the covariant derivative in these cases since the geometry is flat.} 
\begin{equation}
    \langle S_a(x) \mathcal{O}_1\cdots \mathcal{O}_n \rangle_{\mathcal{C}=0} =-2ic_{1,1} \frac{\delta}{\delta \widetilde{\mathcal{C}}^a(x)}\langle \mathcal{O}_1\cdots \mathcal{O}_n  \rangle_{\mathcal{C}}\big{|}_{\mathcal{C}=0} \; ,
\end{equation}
i.e., that the soft-photon insertion enacts parallel transport about the infinite-dimensional vacuum manifold. In this formula, we can view the shadow transform as a change of coordinates. 
Moving the shadow transform around inside the action, one has the equivalent formula for single insertions
\begin{equation}
    \langle J_a(x) \mathcal{O}_1\cdots \mathcal{O}_n \rangle_{\mathcal{C}=0} = -i\frac{\delta}{\delta \mathcal{C}^a(x)}\langle \mathcal{O}_1\cdots \mathcal{O}_n  \rangle_{\mathcal{C}} \big{|}_{\mathcal{C}=0}\; .
\end{equation}
This formula says that $\mathcal{C}_a(x)$ acts as a background gauge field  for the global $U(1)$ current $J_a(x)$.

In the sigma model, changing the boundary condition for the bulk scalar turned on a source for the marginal operator dual to the soft scalar \eqref{eq:transport}.
What is the deformation corresponding to a change in the boundary condition \eqref{eq:abelianBC} in abelian gauge theory? It is simply a current-connection interaction
\begin{equation}\label{eq:CurrentBackground}
   \delta S= \int d^d x \, J^a(x)\mathcal{C}_a(x) \; ,
\end{equation}
i.e., a background gauge field for the global $U(1)$ symmetry of the model.  This statement is familiar in AdS holography. In AdS space, fields have a normalizable mode associated to a dual local operator, as well as a non-normalizable mode whose boundary condition is interpreted as a source for the corresponding local operator. In the case of gauge theory, the non-normalizable mode acts as a background gauge field for the associated current operator in the dual.

As in the sigma model, turning on a coherent state of soft photons parallel transports around the space of vacua:
\begin{equation}\label{eq:MaxwellVac}
    \langle \mathcal{O}_1\cdots \mathcal{O}_n \rangle_{\mathcal{C}_a-\delta \mathcal{C}_a}= \langle\mathcal{O}_1\cdots \mathcal{O}_n e^{\frac{-i}{2c_{1,1}}\int d^d x \, \delta \widetilde{\mathcal{C}}_a(x) S^a(x)}\rangle_{\mathcal{C}_a} \; .
\end{equation}
This obviously resembles the sigma model deformation formula \eqref{eq:transport} if we think of the index $I$ as running over both polarization \textit{and position} in the infinite-dimensional case. 
To determine if this parallel transport is path dependent, we could calculate the curvature of $\mathcal{G}/G$ using the antisymmetric double-soft theorem. It vanishes, and so does the curvature: 
\begin{equation}
 \left[\lim_{q(x)\to 0}\omega,\lim_{q'(y)\to 0}\omega' \right]A_{n+2}^{K_1\cdots K_n;ab }=   R\left(\frac{\delta }{\delta \widetilde{\mathcal{C}}_a(x)}, \frac{\delta }{\delta \widetilde{\mathcal{C}}_b(y)} \right)A_n^{K_1\cdots K_n}= 0 \; .
\end{equation}
The curvature vanishes because the infrared sector is described by a linear model.

Interestingly, the formula \eqref{eq:SoftInt} can also be derived from a slight generalization of formulas that do not directly deal with infrared divergences or soft-exchange. The authors of \cite{He:2020ifr} performed canonical quantization of four-dimensional gauge theory with the boundary conditions \eqref{eq:abelianBC} and found a formula relating $S$-matrix elements in different vacua. The higher dimensional version of their formula (5.8) would be (up to normalization)
\begin{equation}\label{eq:Exp1}
    \langle \mathcal{O}_1\cdots \mathcal{O}_n \rangle_{\mathcal{C}=\partial \varepsilon}= \exp \left[-i\int d^d x\, \partial_a j^a(x,x_k)\,\varepsilon (x) \right]\langle \mathcal{O}_1\cdots \mathcal{O}_n \rangle_{ \mathcal{C}=0} \; , \qquad 
    j^a=\frac{1}{2c_{1,1}} \widetilde{\mathcal{J}}_a \; .
\end{equation}
Integrating by parts, this can equivalently be written
\begin{equation}\label{eq:Exp2}
    \langle \mathcal{O}_1\cdots \mathcal{O}_n \rangle_{\partial \varepsilon}= \exp \left[\frac{i}{2c_{1,1}}\int d^d x \widetilde{\mathcal{J}}^a\mathcal{C}_a \right]\langle \mathcal{O}_1\cdots \mathcal{O}_n \rangle_{ \mathcal{C}=0}
    =\exp \left[\frac{i}{2c_{1,1}}\int d^d x \, \mathcal{J}^a\widetilde{\mathcal{C}}_a \right]\langle \mathcal{O}_1\cdots \mathcal{O}_n \rangle_{ \mathcal{C}=0}
\end{equation}
which is \eqref{eq:SoftInt}. A formula like this for a finite deformation is possible precisely because the space of deformations is flat. Otherwise, there would be path dependence. 

Some authors choose to parameterize the space of vacua in Abelian gauge theory in terms of a field $\varepsilon(x)$, with the corresponding flat connection given by $\mathcal{C}_a=\partial_a\varepsilon$. As we discuss in appendix \ref{app:Derivatives}, there is a simple relationship between derivatives with respect to $\widetilde{\mathcal{C}_a}(x)$ and with respect to $\varepsilon(x)$
\begin{equation}\label{eq:AbelianAltD}
    \frac{\delta}{\delta \widetilde{\mathcal{C}_a}(x)}=\frac{1}{ 2c_{1,1}}\int d^dy\partial_a\log (x-y)^2\frac{\delta}{\delta \varepsilon(y)} \; . 
\end{equation}
This relationship can be seen directly as follows. We write
\begin{equation}
    \partial_a \varepsilon(x)=\mathcal{C}_a(x)=\frac{1}{c_{1,1}}\int d^d y\frac{\mathcal{I}_{ab}(x-y)}{(x-y)^2}\widetilde{\mathcal{C}}_b(y)=\frac{1}{2c_{1,1}}\partial_a \int d^d y \partial_b \log (x-y)^2\widetilde{\mathcal{C}}_b(y) \; .
\end{equation}
Removing the derivatives and taking a functional derivative yields  
\begin{equation}
    \frac{\delta \varepsilon (x)}{\delta \widetilde{\mathcal{C}}_a(y)}=\frac{1}{2c_{1,1}}\partial_a \log(x-y)^2 \; .
\end{equation}
Viewing the shadow transform as an infinite-dimensional change of coordinates, we find
\begin{equation}\label{eq:AbCoords}
    \frac{\delta}{\delta \widetilde{\mathcal{C}_a}(x)}=\int d^dy \frac{\delta \varepsilon (y)}{\delta \widetilde{\mathcal{C}}_a(x)}\frac{\delta}{\delta \varepsilon(y)}=\frac{1}{ 2c_{1,1}}\int d^dy\partial_a\log (x-y)^2\frac{\delta}{\delta \varepsilon(y)} \; . 
\end{equation}
Written this way, it is obvious that \eqref{eq:AbelianAltD} acting on \eqref{eq:Exp1} yields the soft theorem. 
As we will see, it is this form of the derivative on the moduli space that appears most naturally in Yang-Mills theory.

\section{Gravity}\label{sec:gravity}
Finite energy boundary conditions for gravity in asymptotically flat spacetimes allow the leading transverse components of the metric to approach a \textit{flat supertranslation connection} at infinity \cite{Strominger:2013jfa,Kapec:2015vwa}
\begin{equation}\label{eq:gravityBC}
    \langle g_{ab}\rangle=\mathcal{C}_{ab}(x) \; , \qquad \partial_{[a} \mathcal{C}_{b]c}  - \frac{1}{d-1} \delta_{c[a} \partial^d \mathcal{C}_{b]d} = 0 \;.
\end{equation}
This notation indicates that, in an asymptotic expansion about null infinity in flat coordinates, the transverse components of the metric have large-$r$ asymptotics ${g_{ab}(u,r,x^a)\sim r^2\delta_{ab} +r\mathcal{C}_{ab}(x) + O(1)}$, with $\mathcal{C}_{ab}(x)$ a flat supertranslation connection independent of the retarded time $u$. The radiative degrees of freedom in the gravitational field are encoded in the subleading terms in the large-$r$ expansion. Since this degree of freedom $\mathcal{C}_{ab}(x)$ contains no $u$-dependence, it is effectively non-dynamical at the classical level but it plays an important role in the infrared dynamics of the quantum theory.\footnote{In four dimensions the radiation term is not subleading, but the $O(r)$ term can be decomposed into a normalizable radiative piece that depends on $u$ and a non-normalizable boundary mode $\mathcal{C}_{ab}(x)$ obeying \eqref{eq:gravityBC}.}

$\mathcal{C}_{ab}$ is symmetric and traceless, and the flatness condition is equivalent to the requirement that
\begin{equation}\label{eq:STflat}
   \mathcal{C}_{ab}(x)=2(\partial_a \partial_b - \frac{1}{d}\delta_{ab}\partial^c\partial_c)f(x) 
\end{equation}
for some $f:\mathbb{R}^d\to \mathbb{R}$. This boundary condition is invariant under global translations (equivalently, the functions $x^a, 1\pm x^2$ are in the kernel of the differential operator appearing in \eqref{eq:STflat}), but it spontaneously breaks all other diffeomorphisms with non-compact support. The space of allowed boundary conditions is the orbit of the zero connection $\mathcal{C}_{ab}=0$ under the action of the supertranslations $\mathcal{G}$. The moduli space is therefore $\mathcal{M}=\mathcal{G}/\mathbb{R}^{d+2}$ since the translations are unbroken on the vacuum. 

In the quantum theory, $S$-matrix elements carry an extra label corresponding to the boundary condition \eqref{eq:gravityBC}. Since that boundary condition spontaneously breaks the supertranslation symmetry, there is an associated Goldstone boson $C_{ab}(x)$ for the coset.  Under supertranslations
\begin{equation}
    C_{ab}(x) \to C_{ab}(x) + 2(\partial_a \partial_b - \frac{1}{d}\delta_{ab}\partial^c\partial_c)f(x)
\end{equation}
so this degree of freedom behaves like a supertranslation Goldstone mode.

Perturbative $S$-matrix calculations are performed with the boundary condition $\mathcal{C}_{ab}=0$, and the soft graviton theorem says
\begin{equation}
\label{soft-grav-thm}
\langle \mathcal{O}_{ab} ( \omega , x ) \mathcal{O}_1 \cdots \mathcal{O}_n \rangle_{\mathcal{C}=0}  ~~ \stackrel{\omega\to0}{\longrightarrow} ~~   \frac{\kappa}{2\omega} \sum_{k=1}^n  \frac{ p_{k\mu} p_{k\nu} \varepsilon^{\mu\nu}_{ab} (x)  }{ p_k \cdot \hat{q}(x) } \langle  \mathcal{O}_1 \cdots \mathcal{O}_n \rangle_{\mathcal{C}=0}  \; , 
\end{equation}
where $\kappa=32\pi G$. The leading soft graviton operator
\begin{equation}
    S_{ab}(x)=\frac{2}{\kappa}\oint \frac{d\omega}{2\pi i }\mathcal{O}_{ab}(\omega,x) \; 
\end{equation}
has matrix elements 
\begin{equation}
\langle S_{ab}(x) \mathcal{O}_1 \cdots \mathcal{O}_n \rangle_{\mathcal{C}=0}  =  \mathcal{J}_{ab}(x)\, \langle \mathcal{O}_1 \cdots \mathcal{O}_n \rangle_{\mathcal{C}=0} \; ,  
\end{equation}
where
\begin{equation}
\mathcal{J}_{ab}(x) \equiv - \left( \partial_a \partial_b - \frac{1}{d} \delta_{ab} \partial^2 \right) \sum_{k=1}^n \omega_k [ - {\hat p}_k \cdot {\hat q}(x) ]  \log [ - {\hat p}_k \cdot {\hat q}(x) ] \;  .\\
\end{equation}
In the presence of an IR regulator $\mu$ and a hard-soft separation scale $\Lambda$, the soft factorization theorems say
\begin{equation}
    \langle \mathcal{O}_1 \cdots \mathcal{O}_n \rangle_{\mathcal{C}=0}^\mu  = e^{ - \Gamma } \langle \mathcal{O}_1 \cdots \mathcal{O}_n \rangle_{\mathcal{C}=0}^\Lambda \; , \quad  \Gamma=\alpha_{\text{gr}} \int  \frac{d^d x}{(2\pi)^d} [\mathcal{J}_{ab}(x)]^2 \; , \quad \alpha_{\text{gr}}= \frac{\kappa^2}{32 \pi}\int^{\Lambda}_\mu \omega^{d-3}d\omega \; .
\end{equation}
We would like to interpret these formulas in analogy with the sigma model and abelian gauge theory. 
In other words, we anticipate a formula of the schematic form
\begin{equation}
    \langle S_{ab}(x) \mathcal{O}_1\cdots \mathcal{O}_n \rangle \sim \frac{\delta}{\delta \widetilde{\mathcal{C}}^{ab}(x)}\langle \mathcal{O}_1\cdots \mathcal{O}_n  \rangle \; .
\end{equation}
Because the infrared sector of gravity is exactly solvable, we can indeed check this statement. 
The soft action for the Goldstone mode and soft graviton was worked out in \cite{Kapec:2021eug} and takes the form
\begin{equation}
S_{\text{soft}} = \frac{\alpha_{\text{gr}}}{(2\pi)^d} \int d^d x \, S_{ab}(x) S^{ab}(x) + \frac{i}{16 c_{1,2}} \int d^d x \,  \widetilde{C}^{ab}(x)(S_{ab}(x)-\mathcal{J}_{ab})  \; .
\end{equation}
Here $\widetilde{C}^{ab}(x)$ is the shadow transform of the Goldstone mode. This action reproduces all leading soft effects in perturbative gravity in the sense that
\begin{equation}\label{eq:gravAnswer}
\langle S_{a_1b_1}(y_1) \cdots S_{a_m b_m}(y_m) \rangle_{\mathcal{C}=0} =  \mathcal{J}_{a_1b_1}(y_1) \cdots \mathcal{J}_{a_m b_m}(y_m) \exp \left[ - \frac{\alpha_{\text{gr}}}{(2\pi)^d} \int d^d x \, \mathcal{J}_{ab}(x) \mathcal{J}^{ab}(x)  \right] \; .
\end{equation}
When there is a background field this action becomes
\begin{equation}
S_{\text{soft}}[\mathcal{C}] = \frac{\alpha_{\text{gr}}}{(2\pi)^d} \int d^d x \, S_{ab}(x) S^{ab}(x) + \frac{i}{16 c_{1,2}} \int d^d x \,  \widetilde{C}_{ab}(x)(S^{ab}(x)-\mathcal{J}^{ab}) +\widetilde{\mathcal{C}}_{ab}(x)S^{ab}(x) \; .
\end{equation}
This path integral can be done exactly and yields
\begin{equation}
    Z[\mathcal{C}]=\exp \left[ - \frac{\alpha_{\text{gr}}}{(2\pi)^d} \int d^d x \, \mathcal{J}_{ab}(x) \mathcal{J}^{ab}(x)  \right] \exp \left[\frac{-i}{16 c_{1,2}} \int d^d x \, \widetilde{\mathcal{C}}_{ab}(x)\mathcal{J}^{ab}(x) \right] \; .
\end{equation}
Comparing with \eqref{eq:gravAnswer}, this formula makes it clear that
\begin{equation}
    \langle S_{ab}(x) \mathcal{O}_1\cdots \mathcal{O}_n \rangle_{\mathcal{C}=0} =16 i c_{1,2} \frac{\delta}{\delta \widetilde{\mathcal{C}}^{ab}(x)}\langle \mathcal{O}_1\cdots \mathcal{O}_n  \rangle_{\mathcal{C}}\big|_{\mathcal{C}=0} \; ,
\end{equation}
so the soft-graviton insertion enacts infinitesimal parallel transport about the infinite-dimensional vacuum manifold. 
Turning on a coherent state of soft gravitons parallel transports around the space of vacua:
\begin{equation}\label{eq:SuperTVac}
    \langle \mathcal{O}_1\cdots \mathcal{O}_n \rangle_{\mathcal{C}_{ab}-\delta \mathcal{C}_{ab}}= \langle\mathcal{O}_1\cdots \mathcal{O}_n e^{\frac{i}{16c_{1,2}}\int d^d x \, \delta \widetilde{\mathcal{C}}_{ab}(x) S^{ab}(x)}\rangle_{\mathcal{C}_{ab}} \; .
\end{equation}
This resembles the sigma model deformation formula \eqref{eq:transport} if the index $I$ runs over both position and polarization.
The curvature is computed using the antisymmetric double-soft limit, and it vanishes \cite{Klose:2015xoa}
\begin{equation}
 \left[\lim_{q(x)\to 0}\omega,\lim_{q'(y)\to 0}\omega' \right]A_{n+2}^{K_1\cdots K_n;ab,cd}=   R\left(\frac{\delta }{\delta \widetilde{\mathcal{C}}_{ab}(x)}, \frac{\delta }{\delta \widetilde{\mathcal{C}}_{cd}(y)} \right)A_n^{K_1\cdots K_n}= 0 \; .
\end{equation}
We conclude that the space of flat supertranslation connections is itself flat.
This is because the Goldstone degree of freedom is described by a linear model.

As remarked previously, the position-dependent boundary conditions \eqref{eq:abelianBC} and \eqref{eq:gravityBC} in gauge theory and gravity differ significantly from the position-independent boundary condition in the sigma model. As a consequence, the background sources \eqref{eq:MaxwellVac} and \eqref{eq:SuperTVac}
obtained by varying the boundary conditions in gauge theory and gravity are position dependent, while the analogous background source in the sigma model is constant (a coupling constant). 

This is a peculiar feature of celestial CFT which does not arise in garden-variety Euclidean CFT. It is however a crucial aspect of CCFT and cannot be ignored or discarded. Indeed, as we discuss in the conclusion to this paper, it is precisely the existence of this space of deformations which allows CCFT to reproduce non-commuting soft limits.  The deformations \eqref{eq:MaxwellVac} and \eqref{eq:SuperTVac} would seem unnatural, except that they actually arise in the natural world.

The unfamiliar form of \eqref{eq:MaxwellVac} and \eqref{eq:SuperTVac}  is a reflection of the subtle interplay between Lorentz invariance and large gauge symmetry in asymptotically flat space. For gravity in four dimensions, the relevant statement is familiar and could be phrased simply as ``there are an infinite number of copies of $SO(3,1)$ inside the BMS group.'' These different copies of $SO(3,1)$, which amount to different definitions of angular momentum, are related by the action of the supertranslations. In other words, the $SO(3,1)$ subgroup that annihilates the $\mathcal{C}_{ab}=2(\partial_a \partial_b - \frac{1}{2}\delta_{ab}\partial^c\partial_c)f(x) $ vacuum is related to the $SO(3,1)$ subgroup that annihilates the standard vacuum $\mathcal{C}_{ab}=0$ through conjugation by the supertranslation corresponding to $f(x)$.
For a long time this was known as
the ``problem of angular momentum in general relativity,'' but this problem is now understood to be a feature rather than a bug.  Gravity in asymptotically flat space simply has an infinite set of vacua corresponding to different combinations of soft gravitons. These gravitons carry zero energy but non-zero spin, so different copies of $SO(3,1)$ are relevant to different vacua. The deformation \eqref{eq:SuperTVac} appears to break Lorentz (conformal) symmetry, but a different $SO(3,1)$ subgroup is preserved after the deformation.

Since the symmetry structure in abelian gauge theory is almost identical to that of gravity, it is clear that qualitatively similar statements apply to \eqref{eq:MaxwellVac}. Large gauge transformations do not commute with the Lorentz group, so the $SO(3,1)$ subgroups relevant to the vacua $\mathcal{C}_a$ and $\mathcal{C}_a'$ are related by the gauge transformation that brings $\mathcal{C}_a$ to $\mathcal{C}_a'$. \eqref{eq:MaxwellVac} and \eqref{eq:SuperTVac} are not standard marginal deformations, but they are precisely what is required for CCFT to reproduce gravity and gauge theory in asymptotically flat space.

\section{Non-abelian gauge theory}\label{sec:Yang-Mills}
Classical finite energy boundary conditions in Yang-Mills theory allow the gauge field to approach a flat connection at infinity \cite{Strominger:2013lka,He:2020ifr}
\begin{equation}\label{eq:nonabelianBC}
    \langle A_a\rangle=\mathcal{C}_a(x) \; , \qquad \mathcal{C}_a(x)=U\partial_a U^{-1} \; . 
\end{equation}
This boundary condition is invariant under a global right $G$-action $U\to  Ug^{-1}$. It spontaneously breaks all (non-global) gauge transformations with non-compact support. The allowed boundary conditions are just the orbit of the zero connection $\mathcal{C}_a=0$ under the action of $\mathcal{G}=\text{Map}(S^{d},G)$. The moduli space is therefore $\mathcal{M}=\mathcal{G}/G$ since $\mathcal{C}_a(x)$ is invariant under $U\to  Ug^{-1}$.

In the quantum theory, $S$-matrix elements carry an extra label corresponding to the boundary condition \eqref{eq:nonabelianBC}. 
Perturbative $S$-matrix calculations are usually performed with the boundary condition $\mathcal{C}=0$, and the soft gluon theorem says
\begin{equation}\label{eq:YMsoftThm}
\langle O_a^I(\omega,x) \mathcal{O}_1 \cdots \mathcal{O}_n \rangle_{\mathcal{C}=0}  ~~ \stackrel{\o\to0}{\longrightarrow} ~~ \frac{1}{\omega}ig_{\text{YM}} \sum_{k=1}^n\frac{ {\hat p}_k \cdot \varepsilon_a(x) }{ {\hat p}_k \cdot {\hat q}(x)  }T_k^I\; \langle \mathcal{O}_1 \cdots \mathcal{O}_n \rangle_{\mathcal{C}=0} \; ,
\end{equation}
where $g_{\text{YM}}$ is the gauge coupling and $T_k^I$ acts in the representation $R_k$ under which the $k$-th particle transforms. The leading soft gluon operator
\begin{equation}
    S^I_a(x)=\frac{1}{g_{\text{YM}}} \oint \frac{d\omega}{2\pi i}\mathcal{O}^I_a(\omega,x) \; 
\end{equation}
has matrix elements
\begin{equation}\label{eq:YMsoft}
\langle S_a^I(x) \mathcal{O}_1 \cdots \mathcal{O}_n \rangle_{\mathcal{C}=0}  = i \sum_{k=1}^n\partial_a \log(x-x_k)^2\; T_k^I\; \langle \mathcal{O}_1 \cdots \mathcal{O}_n \rangle_{\mathcal{C}=0} \; .
\end{equation}
A straightforward computation shows that the shadow transform $J_a^I(x)=\frac{1}{2c_{1,1}}\widetilde{S}^I_a(x)$ obeys the Ward identity for a non-abelian conserved current
\begin{equation}
\langle \partial^a J_a^I(x) \mathcal{O}_1 \cdots \mathcal{O}_n  \rangle_{\mathcal{C}=0} =  i\sum_{k=1}^n  \delta^{(d)}(x-x_k)T_k^I  \langle  \mathcal{O}_1 \cdots \mathcal{O}_n \rangle_{\mathcal{C}=0} \;  ,
\end{equation}
and symmetry considerations still imply that the boundary condition \eqref{eq:nonabelianBC} should act as a background source for the global symmetry current $J_a^{I}(x)$
\begin{equation}
    \langle J^I_a(x) \mathcal{O}_1\cdots \mathcal{O}_n \rangle_{\mathcal{C}=0} \sim \frac{\delta}{\delta \mathcal{C}_I^a(x)}\langle \mathcal{O}_1\cdots \mathcal{O}_n  \rangle_{\mathcal{C}} \big{|}_{\mathcal{C}=0}\; .
\end{equation}
Following the discussion in abelian gauge theory and gravity, we would like to interpret \eqref{eq:YMsoft} as a functional derivative on the space of vacua
\begin{equation}\label{eq:NAsoftD}
    \langle S^I_{a}(x) \mathcal{O}_1\cdots \mathcal{O}_n \rangle \stackrel{?}{\sim} \frac{D}{D \widetilde{\mathcal{C}}_I^{a}(x)}
    \langle \mathcal{O}_1\cdots \mathcal{O}_n  \rangle \; .
\end{equation}
The notation $\frac{D}{D \widetilde{\mathcal{C}}_{\small I}^{a}(x)}$ denotes a covariant functional derivative and anticipates the likelihood that the space of vacua in Yang-Mills theory is curved, in contrast with abelian gauge theory and gravity.

The $d$-dimensional soft action has not been derived for non-abelian gauge theory. 
However,  it is intuitively clear that a formula of the general form \eqref{eq:NAsoftD} should hold in Yang-Mills theory. Studying the classical phase space of non-abelian gauge theory on non-compact spaces, one finds that the soft mode $S_a^I(x)$ of the radiative degree of freedom and the boundary degree of freedom $\mathcal{C}_a^I$ are coupled in the symplectic form \cite{He:2020ifr,He:2023bvv}. Upon quantization, one acts as a derivative with respect to the other, and since the standard quantization diagonalizes $\mathcal{C}_a^I$, $S_a^I(x)$ therefore acts as a derivative. Indeed, the authors of \cite{He:2020ifr} performed canonical quantization of non-Abelian gauge theory, including a careful treatment of the zero-modes and boundary conditions. For a flat connection $\mathcal{C}=UdU^{-1}$ with $U=\exp[\phi(x)]$, the authors derived a formula which we show in appendix \ref{app:Derivatives} can be expressed
\begin{equation}\label{eq:SYM}
    \langle S_a(x) \cdots \rangle_{\mathcal{C}} = -iU(x)\int d^dy \partial_a \log (x-y)^2U^{-1}(y)\frac{\text{ad}_{\phi(y)}}{1-\exp[-\text{ad}_{\phi(y)}]}\frac{\delta}{\delta \phi(y)}\langle \cdots \rangle_{\mathcal{C}}\;.
\end{equation}
Formula \eqref{eq:SYM} is clearly the non-Abelian version of \eqref{eq:AbelianAltD}, with the integral arising from the switch to a shadow variable and the nonlinear terms in $\phi(x)$ arising due to the curvature of $G$. Indeed, linearizing about $\mathcal{C}=0$, one has $\mathcal{C}_a=\partial_a\phi(x)$ and $S_a(x)=\int d^dy \partial_a \log(x-y)^2 \frac{\delta}{\delta \phi(y)}\sim \frac{\delta}{\delta \widetilde{C}(x)}$, motivating the notation in \eqref{eq:NAsoftD}. Away from the origin, the right hand side of \eqref{eq:SYM} defines the differential operator in \eqref{eq:NAsoftD} and the nonlinear terms are responsible for the noncommutativity of soft limits.

In contrast to abelian gauge theory and gravity, soft limits in Yang-Mills theory do not commute \cite{Berends:1988zn,Catani:1999ss,Larkoski:2014bxa,Klose:2015xoa}, so the parallel transport induced by soft gluon insertions is path dependent. We review this calculation in appendix \ref{app:Double}. The result can be put in a particularly simple form\footnote{For a general set of vector fields $R(X, Y )Z = \nabla_X \nabla_Y Z -\nabla_Y \nabla_X Z- \nabla_{[X,Y ]}Z$. Since the vector fields $X=\frac{\delta}{\delta \mathcal{C}(x)}$ and $Y=\frac{\delta}{\delta \mathcal{C}(y)}$ involve the flat connection at separated points they commute so the third term vanishes.}
\begin{align}\label{eq:gluonAnti}
 \left[\lim_{q_I(x)\to 0} \omega,\lim_{q_J'(y)\to 0} \omega' \right]A_{n+2}^{K_1\cdots K_n; IJ,ab}
 &=-2g_{\text{YM}}^2 \sum_{k=1}^n \frac{\mathcal{I}_{ac}(x-x_k)\mathcal{I}_{cb}(y-x_k)}{(x-y)^2} f^{IJK}T_k^K A_n^{K_1\cdots K_n},
\end{align}
where the kinematic part of the summand is usually written
\begin{equation}
\frac{p_k \cdot \varepsilon_a}{p_k\cdot \hat{q}}\frac{p_k\cdot \varepsilon_b}{p_k\cdot \hat{q}'}
    +\frac{\varepsilon_a\cdot \varepsilon_b}{\hat{q}\cdot \hat{q}'}
    -\frac{\hat{q}\cdot \varepsilon_b}{\hat{q}\cdot \hat{q}'}\frac{p_k \cdot \varepsilon_a}{p_k\cdot \hat{q}}-\frac{\varepsilon_a\cdot \hat{q}'}{\hat{q}\cdot \hat{q}'}\frac{\varepsilon_b\cdot p_k}{p_k\cdot \hat{q}'} \; .
\end{equation}
Formula \eqref{eq:gluonAnti} computes the curvature of the space of vacua at the point $\mathcal{C}=0$ since that is where the perturbative calculations are performed. Turning on a coherent state of soft gluons parallel transports $S$-matrix elements around the space of vacua, but for finite deformations, the analog of formulas \eqref{eq:MaxwellVac} and \eqref{eq:SuperTVac} requires specification of a path and involves a path-ordered exponential.

There are a couple of important points to note about \eqref{eq:gluonAnti}. First, the expression involves a \textit{single} sum (rather than the double sum that would be present in the simultaneous soft limit, whose geometric interpretation is unclear) over the remaining hard external particles, as it should if it represents curvature. In addition, the position independent factor in \eqref{eq:gluonAnti} bears a strong resemblance to familiar formulas for the Riemannian geometry of bi-invariant metrics on compact Lie groups, especially 
when all of the external states in the amplitude transform in the adjoint representation (as they do in pure Yang-Mills).

To see this, note that the position independent factor in \eqref{eq:gluonAnti} can be expressed as $[T^I,T^J]$ acting on the amplitude. When the external states transform in the adjoint, this is equivalently written $\text{ad}_{[T^I,T^J]}$. 
On a compact Lie group with a bi-invariant metric, the covariant derivative of left-invariant vector fields is $\nabla_XY=\frac12[X,Y]$ and the curvature takes the simple form ${R(X,Y)Z=\frac14\text{ad}_{[X,Y]}Z}$. In other words, $\nabla_{T^I}=\frac12 \text{ad}_{T^I}$ and $R(T^I,T^J)=\frac{1}{4} \text{ad}_{[T^I,T^J]}$. It is interesting to compare these finite-dimensional formulas to the Yang-Mills soft theorem
\begin{equation}
\langle S_a^I(x) \mathcal{O}_1 \cdots \mathcal{O}_n \rangle_{\mathcal{C}=0}  = i \sum_{k=1}^n\partial_a \log(x-x_k)^2\; \text{ad}_{T^I}\; \langle \mathcal{O}_1 \cdots \mathcal{O}_n \rangle_{\mathcal{C}=0} \;\quad \text{vs.} \quad\nabla_{T^I}=\frac12 \text{ad}_{T^I}
\end{equation}
and the antisymmetric double soft theorem 
\begin{equation}
    -2g^2 \sum_{k=1}^n \frac{\mathcal{I}_{ac}(x-x_k)\mathcal{I}_{cb}(y-x_k)}{(x-y)^2} \text{ad}_{[T^I,T^J]} A_n^{K_1\cdots K_n} \; \quad \text{vs.} \quad R(T^I,T^J)=\frac{1}{4} \text{ad}_{[T^I,T^J]} \; .
\end{equation}
In other words at the point $\mathcal{C}=0$ we seem to have a formula of the form
\begin{align}
 \left[\lim_{q_I(x)\to 0} \omega,\lim_{q_J'(y)\to 0} \omega' \right]A_{n+2}^{K_1\cdots K_n; IJ,ab}
 &=-8g^2 \sum_{k=1}^n \frac{\mathcal{I}_{ac}(x-x_k)\mathcal{I}_{cb}(y-x_k)}{(x-y)^2} R(T^I,T^J) A_n^{K_1\cdots K_n}.
\end{align}
It is not uncommon for the curvature of a space of maps from $M$ to $N$ to be controlled by the curvature of $N$: in many examples the infinite dimensional curvature is controlled by a finite dimensional curvature.

It would be interesting to find an independent derivation of \eqref{eq:gluonAnti} as the curvature of some ($d$-dimensional) field space whose dynamics computes the soft exchange and soft emission in $(d+2)$-dimensional non-Abelian gauge theory (perhaps along the lines of \cite{Kapec:2021eug}). Indeed, if there really is a 2d model that captures the IR sector of 4d Yang-Mills theory, then \eqref{eq:gluonAnti} is an important clue since it should describe the metric on the space of fluctuations. 

In the absence of such an off-shell description of the geometry, the soft theorems are still sufficient to define the vacuum geometry in Yang-Mills theory.  Invariant properties of quantum field theories are defined in terms of observable quantities like the $S$-matrix and the spectrum, not the scheme dependent couplings in the Lagrangian. If a quantity is meaningful, it can be expressed in these terms, so the multi-soft limits of scattering amplitudes really serve as the definition of the vacuum geometry, even in the nonlinear sigma model example of section \ref{sec:sigma}. Indeed, the metric on the moduli space receives quantum corrections and can even disappear  non-perturbatively (as it does for pure Yang-Mills theory in the confining phase without any soft particles), which is most easily phrased in the soft theorem language.

\section{Discussion and implications for CCFT}\label{sec:Discussion}

We have seen that the leading soft theorems in gauge theory and gravity have natural interpretations in terms of the infinite-dimensional geometry of vacuum manifolds. This point of view has some natural extensions that would be interesting to investigate. For instance, the  subleading soft graviton theorem is universally related to the stress tensor of CCFT through the shadow transform \cite{Kapec:2017gsg}. The corresponding soft operator is believed to be connected to an infinite-dimensional space of superrotation vacua \cite{Barnich:2010eb,Kapec:2014opa}, although there is still some confusion regarding the precise definition of the infinite-dimensional manifold in this case \cite{Campiglia:2014yka}. Consecutive soft limits at this order in the soft expansion do not commute, and it seems likely that the corresponding antisymmetric double-soft limit can be interpreted as curvature and might even be used to determine the correct infinite-dimensional manifold. The path integral on this space, with the metric determined by the antisymmetric double-soft limit, will likely relate to universal stress tensor dynamics in CCFT and might be used to compute subleading virtual soft exchange in gravity. On a related note, several recent works have studied an infinite tower of soft operators \cite{Strominger:2021mtt,Himwich:2021dau,Adamo:2021lrv} in the context of $w_{1+\infty}$ symmetry, and it would be interesting to integrate those developments into this formalism. Recent work \cite{Cheung:2022vnd} on ``geometry-kinematics duality'' expressed tree-level Yang-Mills scattering amplitudes in terms of the curvature of an infinite-dimensional ``kinematic metric,'' and it would be interesting to understand the relationship of that work to the approach adopted here.

It would be a huge advance to finally produce a 2d action that computes soft exchange and soft emission in 4d non-abelian gauge theory. Heuristically, one hopes to relate the long-wavelength fluctuations of the 4d gluon to the theory of non-abelian Goldstone bosons in two dimensions. Both models describe a broken phase in perturbation theory but generate non-perturbative gaps in the spectrum. If one could relate the two models order by order in perturbation theory, it might be possible to exploit information about the non-perturbative behavior in the 2d model to explain the strong dynamics in 4d. After all, it is really the long-wavelength modes in Yang-Mills that are strongly coupled and that might have a simple dual description. 

This program was carried out for abelian gauge theory and gravity in any dimension in \cite{Kapec:2021eug}, but the generalization to the non-abelian case encounters several obstacles. 
First, in the language of this paper, the infinite-dimensional metric that governs the Goldstone dynamics is flat in the cases considered in \cite{Kapec:2021eug} while it is curved for non-abelian gauge theory. In other words, the 2d actions that compute soft exchange and soft emission in abelian gauge theory and gravity are free, while the analogous model for Yang-Mills theory will be interacting. Formula \eqref{eq:gluonAnti} is therefore an important piece of data for the construction, since it represents the curvature of the tree-level metric. It is known that  the soft-limits receive quantum corrections in non-abelian gauge theory, which means that the metric on the moduli space will receive quantum corrections as well.

There is also a longstanding puzzle regarding how to construct an intrinsically two-dimensional model capable of incorporating the non-commutativity of four-dimensional soft limits. Until now, it was difficult to even phrase this question in 2d language, since there is really no notion of ``inserting a \textit{local} operator first'' in a \textit{single} Euclidean CFT$_d$.
Non-commutativity of soft gluon limits is a crucial aspect of the bulk gauge theory, and is a reflection of properties believed to be crucial in CCFT. It is a feature, not a bug, and it should have a natural expression in the celestial dual. 

The results of this paper offer new insight into this question. The key point is that the dual CCFT
\textit{is more than just a $d$-dimensional system with $G$ symmetry.} It also comes with a space of deformations $\mathcal{G}/G$, and that space is curved. This is a highly unusual property for a CFT$_d$ to have, and has not been considered in the literature until now. As unfamiliar as it may seem, we cannot dispense with this aspect of CCFT: the soft behavior of tree-level scattering amplitudes directly implies it, and it offers a natural, intrinsically $d$-dimensional explanation of the non-commutativity of soft gluon limits.

To understand what is going on, it is helpful to revisit the analysis of the sigma model. The anti-symmetric double-soft limit in the sigma model computes the curvature of the CCFT$_d$ conformal manifold \cite{Kapec:2022axw}, which is a quantity definable purely within CFT$_d$. How is a statement like this possible, when local operators in Euclidean CFT have no notion of ``insertion ordering?'' The key point is that
marginal operators are local operators, but it is the shadow transform of the marginal operator that corresponds to the soft scalar. Because they are non-local \textit{integrated deformations,} the soft scalars can have ordering ambiguities that correspond to path dependence of parallel transport on the conformal manifold. While there is no sense in which local operators have an insertion ordering, \textit{it does matter in which order deformations are performed when the space of deformations is curved.}

To be even more concrete, recall that in standard abstract conformal field theory, the curvature is calculated by anti-symmetrizing a doubly-integrated four-point function of marginal operators \cite{Kutasov:1988xb} 
\begin{equation*}
\partial \partial g \sim \int d^dx \int d^dy \,\langle M(x)M(y)M(0)M(1)\rangle \; .
\end{equation*}
The local operators in the correlator commute at non-coincident points, but the curvature is given by an integrated correlator which is generically divergent and has to be regulated. This integrated correlator does not obey the standard axioms of a single CFT, and that is how we ``get operator ordering out of Euclidean CFT.'' Local operator insertions have no ordering, but deformations do. 

Given that the analogy between the the sigma model and Yang-Mills theory holds so precisely, it seems likely that the soft gluon insertion should really be viewed as an integrated operator, i.e., a deformation of CCFT. 
With this interpretation, consecutive insertions do not need to commute the way that local operator insertions do. This suggests that we should \textit{view the current as fundamental and the soft insertion as its shadow.} In other words, we would say that the local operator spectrum of the model contains $J_a^I(x)$ but not $S_a^I(x)$, and that $S_a^I(x)$ instead arises as an integrated deformation that relates different points on the moduli space $\mathcal{G}/G$.
Given that integrated correlation functions can in general be scheme dependent, it seems plausible that the simultaneous soft limits that do not have geometric interpretations correspond to ``scheme dependent'' calculations in CCFT.

The results of this paper point to the existence of a unique CCFT$_d$ capable of reproducing all of the infrared behavior of Yang-Mills scattering amplitudes that naturally incorporates non-commuting soft limits. Constructing this model is a priority.

\section*{Acknowledgements}
We would like to thank Albert Law, Prahar Mitra, Puskar Mondal, Sruthi Narayanan, Monica Pate, Ana-Maria Raclariu and Andrew Strominger for comments on the draft. This work is supported by the Center of Mathematical Sciences and Applications as well as DOE grant de-sc/0007870.

\appendix

\section{Antisymmetric double-soft limit}
\label{app:Double}
In this section we review the derivation of the antisymmetric double-soft gluon theorem \cite{He:2020ifr} and rewrite it in compact form. We will suppress the boundary condition label on $S$-matrix elements $\langle \cdot \rangle_{\mathcal{C}(x)}$ throughout, as well as operator insertions which are not to be taken soft. All matrix elements are calculated in the vacuum $\mathcal{C}(x)=0$ and $g\equiv g_{\text{YM}}$.
The single soft gluon theorem says
\begin{equation}
    \langle O^I_a(\omega,x) \cdots       \rangle \xrightarrow[]{\omega\to 0}ig\left[  \sum_{k=1}^n \frac{p_k \cdot \varepsilon_a(x)}{p_k\cdot q(x)}T_k^I\right] \langle \cdots   \rangle \; .  
\end{equation}
We now consider the consecutive soft limit of an $S$-matrix element involving two gluons. Our convention for terms in the commutator $[\lim_{q\to 0},\lim_{q'\to 0}]$ is that the expressions act consecutively from right to left. Taking one gluon soft yields
\begin{equation}
\langle O^I_a(\omega,x)O^J_b(\omega',y)  \cdots      \rangle \xrightarrow[]{\omega'\to 0} ig\left[ \frac{q(x)\cdot\varepsilon_b(y)}{q(x)\cdot q'(y)}f^{IJK}+ \delta^{IK}\sum_{k'=1}^n \frac{p_{k'}\cdot \varepsilon_b(y)}{p_{k'}\cdot q'(y)}T_{k'}^J  \right] \langle O^K_a(\omega,x)\cdots    \rangle \; ,  
\end{equation}
and taking the second soft limit one finds
\begin{equation}
\langle O^I_a(\omega,x)O^J_b(\omega',y)  \cdots      \rangle \xrightarrow[]{\omega', \; \text{then}\; \omega} (ig)^2\left[ \frac{q(x)\cdot \varepsilon_b(y)}{q(x)\cdot q'(y)}f^{IJK}+ \delta^{IK}\sum_{k'=1}^n \frac{p_{k'}\cdot \varepsilon_b(y)}{p_{k'}\cdot q'(y)}T_{k'}^J  \right] \sum_{k=1}^n\frac{p_{k}\cdot \varepsilon_a(x)}{p_{k}\cdot q(x)}T_{k}^K\langle \cdots    \rangle \; .  
\end{equation}
Taking the limits in the reverse order gives
\begin{equation*}
\langle O^I_a(\omega,x)O^J_b(\omega',y)    \cdots    \rangle \xrightarrow[]{\omega,\; \text{then} \; \omega'} (ig)^2\left[ \frac{q'(y)\cdot \varepsilon_a(x)}{q'(y)\cdot q(x)}f^{JIK}+ \delta^{JK}\sum_{k=1}^n \frac{p_k\cdot \varepsilon_a(x)}{p_k\cdot q(x)}T_k^I  \right] \sum_{k'=1}^n\frac{p_{k'}\cdot \varepsilon_b(y)}{p_{k'}\cdot q'(y)}T_{k'}^K\langle \cdots    \rangle \; .  
\end{equation*}
The commutator involves two types of terms. The first is a double sum that collapses to one sum
\begin{equation}
  (ig)^2\sum_{k'=1}^n \frac{p_{k'}\cdot \varepsilon_b(y)}{p_{k'}\cdot q'(y)}  \sum_{k=1}^n\frac{p_{k}\cdot \varepsilon_a(x)}{p_{k}\cdot q(x)}\left[T^J_{k'}T_{k}^I\right]\langle \cdots    \rangle 
  = (ig)^2\sum_{k=1}^n\frac{p_{k}\cdot \varepsilon_b(y)}{p_{k}\cdot q'(y)}\frac{p_{k}\cdot \varepsilon_a(x)}{p_{k}\cdot q(x)}f^{JIK}T^K_k\langle \cdots    \rangle \;   
\end{equation}
since $[T_k,T_{k'}]=0$ when $k\neq k'$. The remaining terms take the form
\begin{equation}
    (ig)^2\frac{q(x)\cdot\varepsilon_b(y)}{q(x)\cdot q'(y)} \sum_{k=1}^n\frac{p_{k}\cdot \varepsilon_a(x)}{p_{k}\cdot q(x)}f^{IJK}T_{k}^K\langle \cdots    \rangle -
    (ig)^2 \frac{q'(y)\cdot\varepsilon_a(x)}{q'(y)\cdot q(x)}f^{JIK} \sum_{k'=1}^n\frac{p_{k'}\cdot \varepsilon_b(y)}{p_{k'}\cdot q'(y)}T_{k'}^K\langle \cdots    \rangle \;  
\end{equation}
which is simply
\begin{equation}
    (ig)^2 \sum_{k=1}^n\left[\frac{q(x)\cdot\varepsilon_b(y)}{q(x)\cdot q'(y)}\frac{p_{k}\cdot \varepsilon_a(x)}{p_{k}\cdot q(x)} + \frac{q'(y)\cdot\varepsilon_a(x)}{q'(y)\cdot q(x)}\frac{p_{k}\cdot \varepsilon_b(y)}{p_{k}\cdot q'(y)} \right]f^{IJK}T_{k}^K\langle \cdots    \rangle\; .   
\end{equation}
We are free to add a term that vanishes by global color conservation $\sum_k T_k\langle \cdots \rangle = 0$. The choice $g^2\sum_k\frac{\varepsilon_a \cdot \varepsilon_b}{q\cdot q'}f^{IJK}T^K_k \langle \cdots\rangle$ brings the commutator 
\begin{align}
&\left[\lim_{q\to 0}\omega,\lim_{q'\to 0}\omega'\right]\mathcal{A}^{ab,IJ}_{n+2}=\notag\\& g^2\sum_{k=1}^{n}\left[\frac{\hat{p}_{k}\cdot \varepsilon_b(y)}{\hat{p}_{k}\cdot \hat{q}'(y)}\frac{\hat{p}_{k}\cdot \varepsilon_a(x)}{\hat{p}_{k}\cdot \hat{q}(x)} -\frac{\hat{q}(x)\cdot\varepsilon_b(y)}{\hat{q}(x)\cdot \hat{q}'(y)}\frac{\hat{p}_{k}\cdot \varepsilon_a(x)}{\hat{p}_{k}\cdot \hat{q}(x)} -
\frac{\hat{q}'(y)\cdot\varepsilon_a(x)}{\hat{q}'(y)\cdot \hat{q}(x)}\frac{\hat{p}_{k}\cdot \varepsilon_b(y)}{\hat{p}_{k}\cdot \hat{q}'(y)}\right]f^{IJK}T^K_k\mathcal{A}_n  \\
&=-2g^2\sum_{k=1}^{n}\left[
2\frac{(x-x_k)_a(y-x)_b}{(x-x_k)^2}
+2\frac{(x-y)_a(y-x_k)_b}{(y-x_k)^2}
-2\frac{(x-x_k)_a(y-x_k)_b}{(x-x_k)^2(y-x_k)^2}(x-y)^2\right]\frac{f^{IJK}T^K_k}{(x-y)^2}\mathcal{A}_n\notag  
\end{align}
into an interesting factorized form
\begin{equation}
\left[\lim_{q\to 0}\omega,\lim_{q'\to 0}\omega'\right]\mathcal{A}^{ab,IJ}_{n+2}= -2g^2\sum_{k=1}^{n}\frac{\mathcal{I}_{ac}(x-x_k)\mathcal{I}_{cb}(y-x_k)}{(x-y)^2}f^{IJK}T^K_k\mathcal{A}_n \; ,  
\end{equation}
where
\begin{equation}
    \mathcal{I}_{ab}(x)=\delta_{ab} -2\frac{x_ax_b}{x^2} \; .
\end{equation}

\section{Coordinate systems on the space of vacua}
\label{app:Derivatives}
In this appendix we collect formulas relating the derivatives on the spaces of vacua in different coordinate systems. 

\subsection*{Abelian gauge theory and flat $U(1)$ connections}
In abelian gauge theory, the natural coordinates on the moduli space are flat connections $\mathcal{C}_a(x)$ satisfying $\partial_{[a}\mathcal{C}_{b]}=0$. On $\mathbb{R}^d$ this equation can always be solved using a trivial gauge transformation
${\mathcal{C}_a(x)=\partial_a\varepsilon(x)}$, but the coordinate $\varepsilon(x)$ must then be identified under constant shifts $\varepsilon(x) \sim \varepsilon (x)+c$ since these transformations leave the flat connection invariant. This zero mode decouples inside of $S$-matrix elements which conserve global electric charge and therefore does not appear even when we express quantities in terms of $\varepsilon(x)$ rather than $\mathcal{C}_a$. Many treatments in the literature prefer to work directly with $\varepsilon(x)$ (which is subject to an identification) rather than $\mathcal{C}_a(x)$ (which is subject to a constraint). In order to compare to previous work, in this appendix we derive the relationships between functional derivatives with respect to $\widetilde{\mathcal{C}}_a$ and with respect to $\varepsilon(x)$. The functional $\mathcal{C}_a[\varepsilon](x)=\int d^d y \delta^{(d)}(x-y) \partial_a \varepsilon(y)$
has the variation
\begin{equation}
    \delta \mathcal{C}_a[\varepsilon](x)= \int d^dy \delta^{(d)}(x-y)\partial_a\delta \varepsilon (y)=\int d^dy \partial_a\delta^{(d)}(x-y)\delta \varepsilon (y)
    \equiv \int d^dy \frac{\delta \mathcal{C}_a[\varepsilon](x)}{\delta \varepsilon(y)}\delta \varepsilon (y)\; ,
\end{equation}
from which we conclude
\begin{equation}\label{eq:ddC}
\frac{\delta \mathcal{C}_a[\varepsilon](x)}{\delta \varepsilon(y)}=\partial_a \delta^{(d)}(x-y) \; .
\end{equation}
Similarly, for a functional $F[\mathcal{C}]$ expressed in terms of $\mathcal{C}_a$
\begin{equation}
  \delta F[\mathcal{C}] = \int d^dx\frac{\delta F[\mathcal{C}]}{\delta \mathcal{C}_a (x)} \partial_a \delta \varepsilon(x) 
  =-\int d^dx\partial_a\frac{\delta F[\mathcal{C}]}{\delta \mathcal{C}_a(x) } \delta \varepsilon(x) 
  \equiv \int d^dx \frac{\delta F[\mathcal{C}]}{\delta \varepsilon (x)}\delta \varepsilon(x) \; ,
\end{equation}
from which we conclude
\begin{equation}\label{eq:ddEp}
    \frac{\delta}{\delta \varepsilon(y)}=-\partial^a \frac{\delta}{\delta \mathcal{C}_a(y)} \; .
\end{equation}
We also have 
\begin{equation}
    \delta^{(d)}(x-y)=\frac{\delta \varepsilon(x)}{\delta \varepsilon(y)}=\int d^d z \frac{\delta \varepsilon(x)}{\delta \mathcal{C}_a(z)}\frac{\delta \mathcal{C}_a(z)}{\delta\varepsilon(y)}
\end{equation}
which combined with \eqref{eq:ddC} allows us to deduce
\begin{equation}
   -\partial^a \frac{\delta \varepsilon(x)}{\delta \mathcal{C}_a(y)}=\delta^{(d)}(x-y) \; .
\end{equation}
According to \eqref{eq:abelianSoft}-\eqref{eq:AbelianWI} this can be solved by
\begin{equation}
    \frac{\delta \varepsilon(x)}{\delta \mathcal{C}_a(y)}=-\frac{1}{2c_{1,1}}\int d^dz \frac{\mathcal{I}_{ab}(x-z)}{[(x-z)^2]^{d-1}}\partial_b \log(z-y)^2 \; ,
\end{equation}
i.e. the shadow transform of the derivative of a logarithm.

The derivatives that appear in the soft theorems are actually derivatives with respect to the shadow transform of the flat connection.
Using the definition of the shadow transform one has
\begin{equation}
    \widetilde{\mathcal{C}}_a[\varepsilon](x)=\int d^d y\frac{\mathcal{I}_{ab}(x-y)}{[(x-y)^2]^{d-1}}\partial^b\varepsilon(y)=-\int d^d y\partial^b\left[\frac{\mathcal{I}_{ab}(x-y)}{[(x-y)^2]^{d-1}}\right]\varepsilon(y) \; . 
\end{equation}
Taking a derivative yields
\begin{equation}\label{eq:dSdphi}
    \frac{\delta \widetilde{\mathcal{C}}_a[\varepsilon](x)}{\delta\varepsilon(y)}=-\partial^b\left[\frac{\mathcal{I}_{ab}(x-y)}{[(x-y)^2]^{d-1}}\right] \; . 
\end{equation}
What one really wants to calculate is the quantity
\begin{equation}\label{eq:ddS}
    \frac{\delta}{\delta \widetilde{\mathcal{C}}_a(x)}=\int d^dy\frac{\delta\varepsilon(y)}{\delta \widetilde{\mathcal{C}_a}(x)}\frac{\delta}{\delta \varepsilon(y)} \; . 
\end{equation}
To evaluate this quantity we write
\begin{equation}\label{eq:epsilonC}
    \partial_a \varepsilon(x)=\mathcal{C}_a(x)=\frac{1}{c_{1,1}}\int d^d y\frac{\mathcal{I}_{ab}(x-y)}{(x-y)^2}\widetilde{\mathcal{C}}_b(y)=\frac{1}{2c_{1,1}}\partial_a \int d^d y \partial_b \log (x-y)^2\widetilde{\mathcal{C}}_b(y) \; .
\end{equation}
Removing the derivatives and taking a functional derivative yields  
\begin{equation}\label{eq:AbelianJacobian}
    \frac{\delta \varepsilon (x)}{\delta \widetilde{\mathcal{C}}_a(y)}=\frac{1}{2c_{1,1}}\partial_a \log(x-y)^2 \; .
\end{equation}
 Plugging this into \eqref{eq:ddS} one finds
\begin{equation}\label{eq:AbelainDShad}
    \frac{\delta}{\delta \widetilde{\mathcal{C}}_a(x)}=\frac{1}{ 2c_{1,1}}\int d^dy\partial_a\log (x-y)^2\frac{\delta}{\delta \varepsilon(y)} \; . 
\end{equation}
This is the covariant version of the abelianization of formula (5.20) of \cite{He:2020ifr} generalized to all dimensions.
It is instructive to check that both sides of \eqref{eq:AbelainDShad} act identically on the exponentiated soft factor
\eqref{eq:Exp1} and \eqref{eq:Exp2}. First, note that acting on \eqref{eq:Exp2} with the left hand side of \eqref{eq:AbelainDShad} gives
\begin{equation}\label{eq:LeftAct}
    \frac{\delta}{\delta \widetilde{\mathcal{C}}_a(x)}\exp \left[\frac{i}{2c_{1,1}}\int d^d y \, \mathcal{J}^b(y)\widetilde{\mathcal{C}}_b(y) \right]=\frac{i}{2c_{1,1}}\mathcal{J}_a(x)\exp \left[\frac{i}{2c_{1,1}}\int d^d y \, \mathcal{J}^b(y)\widetilde{\mathcal{C}}_b(y) \right]\; .
\end{equation}
Acting with the right hand side of \eqref{eq:AbelainDShad} on \eqref{eq:Exp1} similarly yields
\begin{align}
    &\frac{1}{2c_{1,1}}\int d^dy\partial_a\log (x-y)^2\frac{\delta}{\delta \varepsilon(y)}\exp \left[-\frac{i}{2c_{1,1}}\int d^d z\, \partial_b \widetilde{\mathcal{J}^b}(z)\,\varepsilon (z) \right] \\
    &=-\frac{i}{4c_{1,1}^2}\int d^dy\partial_a\log (x-y)^2\partial_b \widetilde{\mathcal{J}^b}(y)\exp \left[-\frac{i}{2c_{1,1}}\int d^d z\, \partial_b \widetilde{\mathcal{J}^b}(z)\,\varepsilon (z) \right] \; .
\end{align}
Integrating by parts performs an additional shadow transform on $\widetilde{\mathcal{J}}$ and we get
\begin{equation}
    \frac{i}{2c_{1,1}} \frac{\widetilde{\widetilde{\mathcal{J}_a}}(x)}{c_{1,1}}\exp \left[-\frac{i}{2c_{1,1}}\int d^d z\, \partial_b \widetilde{\mathcal{J}^b}(z)\,\varepsilon (z) \right] =\frac{i}{2c_{1,1}}\mathcal{J}_a(x)\exp \left[-\frac{i}{2c_{1,1}}\int d^d z\, \partial_b \widetilde{\mathcal{J}^b}(z)\,\varepsilon (z) \right] 
\end{equation}
in agreement with \eqref{eq:LeftAct}.
There is another way to see why \eqref{eq:AbelainDShad} should be proportional to the derivative of a logarithm. Using \eqref{eq:dSdphi}
we write
\begin{equation}
    \frac{\delta \varepsilon(x)}{\delta \varepsilon(y)}=\delta^{(d)}(x-y)
    =\delta_{ab}\int d^d z\frac{\delta \varepsilon(x)}{\delta \widetilde{\mathcal{C}_a}(z)}\frac{\delta \widetilde{\mathcal{C}^b}(z)}{\delta \varepsilon(y)}
    =-\delta_{ab}\int d^d z\frac{\delta \varepsilon(x)}{\delta \widetilde{\mathcal{C}_a}(z)}\partial^c\left[\frac{\mathcal{I}_{bc}(z-y)}{[(z-y)^2]^{d-1}}\right] \; . 
\end{equation} 
In order to solve for $\frac{\delta \varepsilon(x)}{\delta \widetilde{\mathcal{C}_a}(z)}$, note that this formula for the delta function is simply the formula for a repeated shadow transform of a $(\Delta,s)=(1,1)$ operator if we choose
\begin{equation}\label{eq:cov}
    \frac{\delta\varepsilon(x)}{\delta \widetilde{\mathcal{C}_a}(y)}\sim\frac{1}{ 2c_{1,1}}\partial_a \log(x-y)^2 \; 
\end{equation}
and integrate by parts.

\subsection*{Non-abelian gauge theory}
Section \ref{sec:Yang-Mills} discussed the vacuum structure of Yang-Mills theory in terms of the flat $G$ connection $\mathcal{C}_a[U](x)=U(x)\partial_a U^{-1}(x)$. The analog of working with the coordinate $\varepsilon(x)$ in the preceding appendix is to write $U(x)=\exp [\phi(x)]$ and to work directly with the Lie algebra valued coordinate $\phi(x)$. Since $\mathcal{C}_a[U](x)$ is invariant under the transformation $U\to Ug^{-1}$ for constant $g\in G$, $\phi(x)$ is subject to an identification resulting from global color conservation.

In \cite{He:2020ifr} it was shown that insertions of the soft gluon in an arbitrary $\mathcal{C}$ vacuum are given by
\begin{equation}\label{eq:IntDeriv}
    \langle S_a(x) \cdots \rangle_{\mathcal{C}} = -iU(x)\int d^dy \partial_a \log (x-y)^2U^{-1}(y)\sum_{n=0}^{\infty}\frac{B_n^+}{n!}\text{ad}^n_{\phi(y)}\frac{\delta}{\delta \phi(y)}\langle \cdots \rangle_{\mathcal{C}}\;,
\end{equation}
where $B_n$ are the Bernoulli numbers defined by the generating function
\begin{equation}
    \frac{z}{e^z-1}=\sum_{m=0}^\infty \frac{B_m^-}{m!}z^m \;, \qquad 
    \frac{z}{1-e^{-z}}=\sum_{m=0}^\infty \frac{B_m^+}{m!}z^m \; . 
\end{equation}

In order to understand this formula, it is helpful to first treat the finite dimensional case.
The matrix exponential $\text{exp}:\mathfrak{g}\to G$ has a differential ${\text{dexp}: T_X\mathfrak{g}\to T_{e^X}}G$ that can be expressed in terms of the differential of left multiplication by $e^X$ at the identity of $G$ \cite{duistermaat2012lie}
\begin{equation}\label{eq:dexpL}
    \text{d}\exp_X=\text{d}L[e^X]_1\circ \;\frac{1-e^{-\text{ad}_X}}{\text{ad}_X} 
    =e^X\frac{1-e^{-\text{ad}_X}}{\text{ad}_X} \; ,
\end{equation}
or in terms of the differential of right multiplication
\begin{equation}\label{eq:dexpR}
    \text{d}\exp_X=\text{d}R[e^X]_1\circ \;\frac{e^{\text{ad}_X}-1}{\text{ad}_X}  \; .
\end{equation}
Since the matrix logarithm is the inverse of the matrix exponential in a neighborhood of the identity, its tangent map $\text{d}\log: T_{e^X}G\to T_X\mathfrak{g}$ is the inverse of \eqref{eq:dexpL}: $\text{d}\log_{e^X}\circ \,\text{d}\exp_X=1$. Therefore
 \begin{equation}
     \text{d}\log_{e^X}=\frac{\text{ad}_X}{1-e^{-\text{ad}_X}}e^{-X}
     \equiv e^{-X}\sum_{n=0}^{\infty}\frac{B^+_n}{n!}\text{ad}^n_{X} \; .
 \end{equation}
Moving to the infinite dimensional case, if $\frac{\delta}{\delta \phi(y)}\in T_{\exp[\phi(y)]}\mathcal{G}$ then 
\begin{equation}
\text{d}\log_{e^{\phi(y)}}\frac{\delta}{\delta \phi(y)}=\frac{\text{ad}_{\phi(y)}}{1-e^{-\text{ad}_{\phi(y)}}}e^{-\phi(y)}\frac{\delta}{\delta \phi(y)}= e^{-{\phi(y)}}\sum_{n=0}^{\infty}\frac{B^+_n}{n!}\text{ad}^n_{\phi(y)} \; .
\end{equation}
So we see that \eqref{eq:IntDeriv} can equivalently be written in a more geometric form
 \begin{equation}
    \langle S_a(x) \cdots \rangle_{\mathcal{C}} = -iU(x)\int d^dy \partial_a \log (x-y)^2\text{d}\log_{\exp[\phi(y)]}\frac{\delta}{\delta \phi(y)}\langle \cdots \rangle_{\mathcal{C}}\;.
\end{equation}
 Of course at the identity, the differential of the exponential and logarithm are the identity, so that
 \begin{equation}
    \langle S_a(x) \cdots \rangle_{\mathcal{C}=0} = -i\int d^dy \partial_a \log (x-y)^2\frac{\delta}{\delta \phi(y)}\langle \cdots \rangle_{\mathcal{C}=0}\;.
\end{equation}
So if we work to linear order in $\phi(x)$, $\mathcal{C}^I_a=\partial_a\phi^I(x)$ and the formulas are the same as in the Abelian case, with matrix indices, and the soft insertion is the derivative with respect to the shadow of the linearized flat connection. Of course in order to compute multiple soft limits one needs to retain the nonlinear pieces.

Unfortunately the Jacobian for the change of variables between $\phi(x)$ and $\widetilde{\mathcal{C}}(x)$ is significantly more complicated than \eqref{eq:AbelianJacobian}. The directional derivative of the matrix exponential can be written
\begin{equation}
    e^\phi\frac{d}{dt} e^{-\phi + tT^I}\big{|}_{t=0}=\frac{\exp[\text{ad}_\phi]-1}{\text{ad}_\phi}T^I \; ,
\end{equation}
so defining
\begin{equation}\label{eq:dlog}
     M[\phi]\equiv\frac{\text{ad}_\phi}{\exp[\text{ad}_\phi]-1}\equiv\sum \frac{B^-_n}{n!}\text{ad}_\phi^n\; 
\end{equation}
allows us to write
\begin{equation}
    \mathcal{C}_a=e^{\phi(x)}\partial_a e^{-\phi(x)}=\frac{\exp[\text{ad}_{\phi(x)}]-1}{\text{ad}_{\phi(x)}}\partial_a \phi(x)=M[\phi]^{-1}\partial_a \phi \;.
\end{equation}
The nonlinear dependence on $\phi(x)$ prevents removing the derivative as in  \eqref{eq:epsilonC}
\begin{equation}
\partial_a\phi(x)=M[\phi(x)]\mathcal{C}_a(x)=M[\phi(x)]\frac{1}{2c_{1,1}}\partial_a\int d^dy \partial_b \log (x-y)^2 \widetilde{\mathcal{C}}_b(y) \;.
\end{equation}

\bibliographystyle{apsrev4-1long}
\bibliography{Bib.bib}
\end{document}